%% file: main.tex
\documentclass[a4paper,11pt]{article}
\usepackage{pos}
\usepackage{siunitx}
\usepackage{todonotes}
\usepackage{cleveref}
\crefname{figure}{Fig.}{Figs.}
\crefname{section}{Section}{Sections}
\usepackage{wrapfig}
\usepackage{enumitem}

\title{Graph Neural Networks for Photon Searches with the Underground Muon Detector of the Pierre Auger Observatory}
 \ShortTitle{GNNs for Photon Searches with the UMD of the Pierre Auger Observatory}

\author*[ab]{Ezequiel Rodriguez}

\affiliation[a]{Instituto de Tecnologías en Detección y Astropartículas (CNEA, CONICET, UNSAM),
Av. General Paz 1555 (B1630KNA) San Martín, Buenos Aires, Argentina}
\affiliation[b]{Institute for Astroparticle Physics (IAP), Karlsruhe Institute of Technology,
P.O. Box 3640, 76021 Karlsruhe, Germany}

\onbehalf{on behalf of the Pierre Auger Collaboration$^c$}
\affiliation[c]{Observatorio Pierre Auger, Av.\ San Mart{\'\i}n Norte 304, 5613 Malarg\"ue, Argentina\\
Full author list: {\rm\url{https://www.auger.org/archive/authors_icrc_2025.html}}}

\emailAdd{spokespersons@auger.org}

\abstract{Ultra-high-energy photons have long been sought as tracers of the most energetic processes in the Universe. Several sources can contribute to a diffuse photon flux, including interactions of cosmic rays with Galactic matter and radiation fields, as well as more exotic scenarios such as the decay of super-heavy dark matter. Regardless of their origin, the expected flux is extremely low, making direct detection impractical and thereby requiring indirect detection by extensive ground-based detector arrays. In this contribution, we present a novel method for photon–hadron discrimination in the energy range of 50 to $\SI{300}{PeV}$ based on deep learning algorithms. Our approach relies on information from both the Surface Detector (SD) and the Underground Muon Detector (UMD) of the Pierre Auger Observatory. The SD consists of an array of water-Cherenkov detectors. It is used to measure the electromagnetic and muonic components of extensive air showers at ground level. Meanwhile, the UMD is composed of buried scintillator modules. It is sensitive to air-shower muons with energies above ${\sim}\SI{1}{\giga \electronvolt}$, enhancing the identification of muon-poor air showers as initiated by photon primaries. Our method represents air-shower events as graphs, and consequently, the network architecture is composed of graph attention layers. We assess the performance of the method on a data subset and discuss the implications of unblinding the full current dataset, as well as the prospects of the increasing data volume expected in the coming years, particularly in terms of sensitivity to various diffuse fluxes from theoretical predictions.}

\ConferenceLogo{PoS_ICRC2025_logo}

\FullConference{39th International Cosmic Ray Conference (ICRC2025)\\
 15–24 July 2025\\
Geneva, Switzerland\\}

\begin{document}
\maketitle

\section{Introduction}
\label{sec:intro}

Ultra-high-energy (UHE) photons serve as valuable tracers of the most energetic phenomena in the Universe. The detection of UHE photons gives insights into cosmic-ray interactions with matter and radiation fields, such as those in the Galactic plane or nearby extragalactic environments, as well as into more exotic scenarios like the decay of super-heavy dark matter (SHDM) particles~\cite{Kampert2011,Bobrikova2021,Berat2022,Abreu2023}. UHE photon fluxes at Earth are expected to originate primarily from the Milky Way and its satellite galaxies. More distant sources are strongly attenuated, as photon interaction lengths are on the order of $0.1\,$Mpc. Despite the rich physics potential, the diffuse photon flux is expected to lie several orders of magnitude below the flux of charged cosmic rays, making direct detection unfeasible. As an alternative approach, large surface detector arrays enable indirect detection by measuring secondary particles from extensive air showers (EAS). In particular, photon-induced showers develop deeper in the atmosphere and produce significantly fewer muons than hadron-initiated showers.

The Pierre Auger Observatory is based on a multi-hybrid design, aimed to study ultra-high-energy cosmic rays using complementary detection techniques. The Surface Detector (SD) comprises an array of water-Cherenkov detectors (WCDs) arranged in triangular grids at 1500, 750, and 433\,m spacings, and measures the electromagnetic, the hadronic, and the muonic components of the EAS. To improve its sensitivity to the different air-shower components, which in turn would provide valuable information toward the discrimination of photon primaries in the cosmic-ray background, the final phase of the AugerPrime upgrade is currently undergoing~\cite{RefPrime}. This includes the installation of the Underground Muon Detector (UMD)~\cite{deJesus2023,perez2024}, composed of buried scintillator modules located next to WCDs in the SD-750 and SD-433. These stations are shielded by 2.3\,m  of soil to suppress the electromagnetic component~\cite{scornaEPJ2023, protoUMD2016} and measure air-shower muons with ${\gtrsim}1\,$GeV. Muons penetrating the soil produce light in the scintillator bars, which is read out by silicon photomultipliers (SiPMs).

In this work, we propose a novel method for photon–hadron discrimination in the energy range from 50 to 300\,PeV using Graph Neural Networks (GNNs). By representing air showers as graphs, we leverage GNNs to process the geometric detector layout and integrate input features from both the SD-433 and UMD. The model architecture is built from graph attention layers, which naturally account for any irregularity in the detector grid. After training and evaluating our method with simulated air-shower events, we apply it to a subset of data used in previous photon search studies~\cite{JCAP2025PhotonHadron}, confirming its future application in larger datasets.

\section{Simulations}
\label{sim_set}

Extensive air showers were generated with \textsc{CORSIKA}~\cite{RefCORSIKA} using the hadronic interaction model EPOS LHC~\cite{RefEPOS}. The detector responses of the SD-433 and the UMD were modeled with the AugerOffline framework~\cite{RefOffline}. Primary particles were sampled isotropically, with uniform distributions in $\sin^2\theta_{\text{MC}}$ and azimuthal angle $\phi_{\text{MC}}$, and energies spanning the range $\lg(E_{\text{MC}}/\mathrm{eV}) \in [16.0, 17.5]$. The simulation includes the arrival of secondary particles at the SD stations and the propagation of muons through 2.3\,m of soil to the UMD, along with the associated optoelectronics response and signal acquisition for each detector.

A preliminary set of photon simulations was used for two purposes: first, to estimate the array photon detection efficiency as a function of the primary energy and zenith angle; and second, to optimize the event reconstruction for photon-initiated showers.

The main set of simulated events was divided into training, validation, and testing data sets. All events in these sets verify a minimum efficiency of 0.9. The training data set includes ${\sim}160,000$ simulated events, equally divided between photon- and proton-induced showers. The validation data set contains ${\sim}40,000$ events with the same composition. The test data set comprises approximately ${\sim}60,000$ photon and proton events, and an additional ${\sim}50,000$ iron-induced showers, allowing assessment of the model's response to a broader range of primary masses. 

\section{Event reconstruction}
\label{sec:photon_rec}

\begin{wrapfigure}{r}{0.5\textwidth}  
    \centering
    \vspace{-10pt} 
    \includegraphics[width=0.5\textwidth]{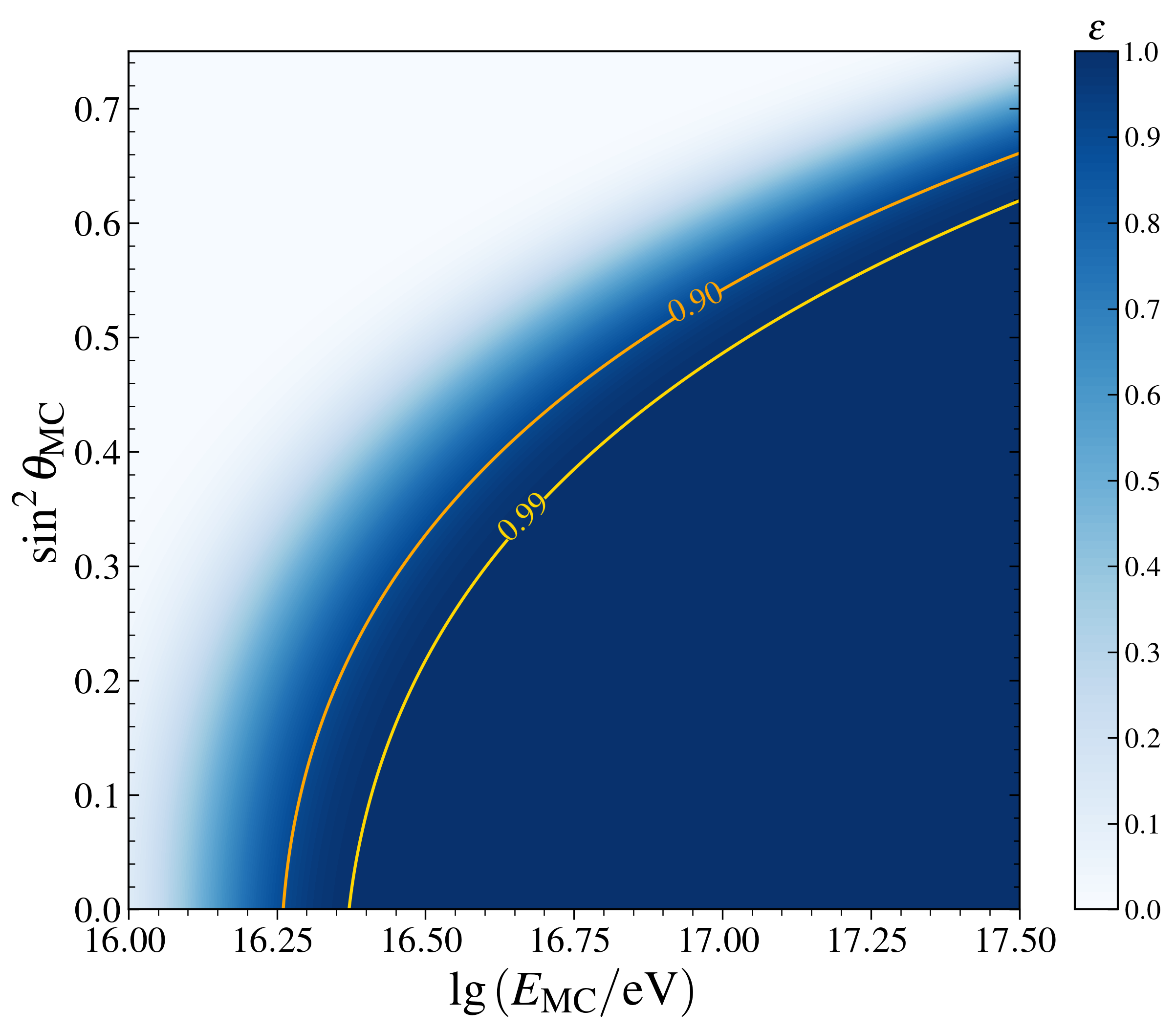}
    \caption{Estimated reconstruction efficiency, $\varepsilon$, for photon-induced showers as a function of $\lg(E_{\text{MC}}/\mathrm{eV})$ and $\sin^2\theta_{\text{MC}}$. Contours for 99\% and 90\% efficiency are shown in yellow and orange, respectively.}
    \label{fig:eff_map}
    \vspace{-10pt} 
\end{wrapfigure}

To quantify the performance of the SD-433 array in detecting photon-induced air showers, we estimated the reconstruction efficiency by fitting a probit model to the photon simulations using a binomial likelihood. The efficiency $\varepsilon(E_{\text{MC}}, \theta)$ quantifies the probability that a photon primary with energy $E_{\text{MC}}$ and zenith angle $\theta_{\text{MC}}$ is successfully reconstructed. In this work, the analysis is restricted to events with $\lg(E_{\text{MC}}/\mathrm{eV}) \geq 16.7$, corresponding to a phase space where photon–hadron discrimination is currently acceptable. The zenith angle is indirectly restricted by the efficiency requirement. The resulting efficiency curves (see \cref{fig:eff_map}) reveal that the SD-433 array remains efficient down to lower energies for vertical showers. This demonstrates the potential for extending photon searches to lower energy regimes in the future, as long as separation remains feasible.

To ensure a consistent energy scale for all primary particles, all events in this study were reconstructed using photon priors. To reconstruct the \emph{photon energy} $E_{\upgamma}$ of a primary, we adopt a dedicated Monte Carlo calibration based on the lateral distribution function (LDF) of the SD signals, similar to~\cite{JCAP2025PhotonHadron}. The energy estimate is based on the expected signal $S(300)$ at 300\,m from the shower axis, selected as the reference distance for the SD-433 due to its optimal balance between resolution and systematics~\cite{brichetto2023sd433}.

It is important to note that the reconstruction procedure assumes the photon hypothesis: the LDF and the inferred energy correspond to the expectation for photon-induced showers. When applied to hadronic primaries, such as protons or iron nuclei, this introduces a bias due to their higher muon content and distinct development. In other words, the approximation $E_{\text{MC}}{\sim}E_{\upgamma}$ holds only for photon events. As a result, both the efficiency estimation and the reconstructed energy of hadronic events (in data or simulations) will be systematically biased, since selection cuts are based on signal observables calibrated for photons. Nevertheless, this does not affect the classification performance, as the deep learning model developed in this work does not use the reconstructed energy as an input feature. The use of a photon energy scale, $E_{\upgamma}$, thus provides a consistent reference for comparing and analyzing all primary types in a photon search.

The energy estimator is defined by
\begin{equation}
\frac{S(300)}{g(\theta)} = \left( \frac{E_{\text{MC}}}{100~\text{PeV}} \right)^{\alpha(\theta)},
\end{equation}
where the power is a third-degree polynomial 
\begin{equation}
\alpha(\theta) = \alpha_0 + \alpha_1 (\cos^2 \theta) + \alpha_2 (\cos^2 \theta)^2 + \alpha_3 (\cos^2 \theta)^3,
\end{equation}
and the attenuation function takes the form of a Gaisser-Hillas curve
\begin{equation}
g(\theta) = g_0 \left( 1 + \frac{x - g_2}{g_1} \right)^{g_1/g_3} \exp\left( -\frac{x - g_2}{g_3} \right), \quad \text{with } x = \sec\theta - \sec 25^\circ.
\end{equation}

Calibration is performed using simulated photon showers that pass an efficiency cut of $\varepsilon > 0.9$. The resulting energy estimator is unbiased to within ${\sim}1\%$ across the interval $\lg(E_{\text{MC}}/\mathrm{eV}) \in [16.4, 17.5]$, with an energy resolution of ${\sim}14\%$ for zenith angles up to $45^\circ$. In the range between $45^\circ$ and $55^\circ$, the bias can increase up to ${\sim}5\%$, and the resolution degrades to ${\sim}30\%$. The attenuation function $g(\theta)$ is characterized by best-fit parameters $g_0 = 23.04 \pm 0.06$, $g_1 = 0.249 \pm 0.014$, $g_2 = -0.076 \pm 0.003$, and $g_3 = 0.165 \pm 0.004$. The exponent $\alpha(\theta)$ is modeled as a cubic polynomial in $\cos^2\theta$, with coefficients $\alpha_0 = 1.96 \pm 0.09$, $\alpha_1 = -4.49 \pm 0.22$, $\alpha_2 = 6.06 \pm 0.09$, and $\alpha_3 = -2.49 \pm 0.06$.

\section{Graph representation and model architecture}
\label{NNs}

An earlier version of the neural network used in this work was presented in~\cite{rodriguez2025neural}, where a dual-graph representation was used to separately encode information from the SD and the combined SD–UMD event. In the current implementation, we adopt a refined design that reduces memory usage, avoids redundancy, and allows for more flexible input handling, while retaining the original concept of separating and analyzing the contributions from the SD, UMD, and their joint information.

Each event is represented as a single undirected graph \(\mathcal{G} = (\mathcal{V}, \mathcal{E})\), where the set of nodes $\mathcal{V}$ correspond to triggered SD stations connected by a set of edges $\mathcal{E}$. The graph is described by three matrices: the adjacency matrix \(A \in \mathbb{R}^{N \times N}\), where \(A_{ij} = 1\) if stations \(i\) and \(j\) are connected (up to second neighbors in the 433\,m triangular grid); the feature matrix \(X \in \mathbb{R}^{N \times F}\), containing for each node spatial and temporal offsets relative to the station with the largest signal, number of working photomultiplier tubes (PMTs) in each WCD, WCD measured signal, muon density \(\rho_\upmu\), and UMD effective area\footnote{The effective area represents the projected area of the UMD modules in the shower front, accounting for the orientation and geometric alignment of the modules relative to the incoming air shower. The muon density \(\rho_\upmu\) is the estimated number of muons divided by such area.}; and the trace matrix \(X_{\text{traces}} \in \mathbb{R}^{N \times T}\), consisting of the first 60 time bins (25\,ns each) of the averaged SD trace over available PMTs. 

Due to data augmentation in simulations or the presence of non-operational UMD stations in data, some nodes may have missing values in UMD-related features. These are handled using precomputed binary masks indicating the validity of each detector component at every station. A dedicated subcomponent, the \emph{Subgraph Extractor}, uses these masks to dynamically isolate the SD-only, UMD-only, and joint SD–UMD subgraphs during training and inference. This approach eliminates the need to store redundant graph copies and significantly reduces memory consumption per event compared to the earlier design~\cite{rodriguez2025neural}.

\begin{figure}
    \centering
    \includegraphics[width=0.98\textwidth]{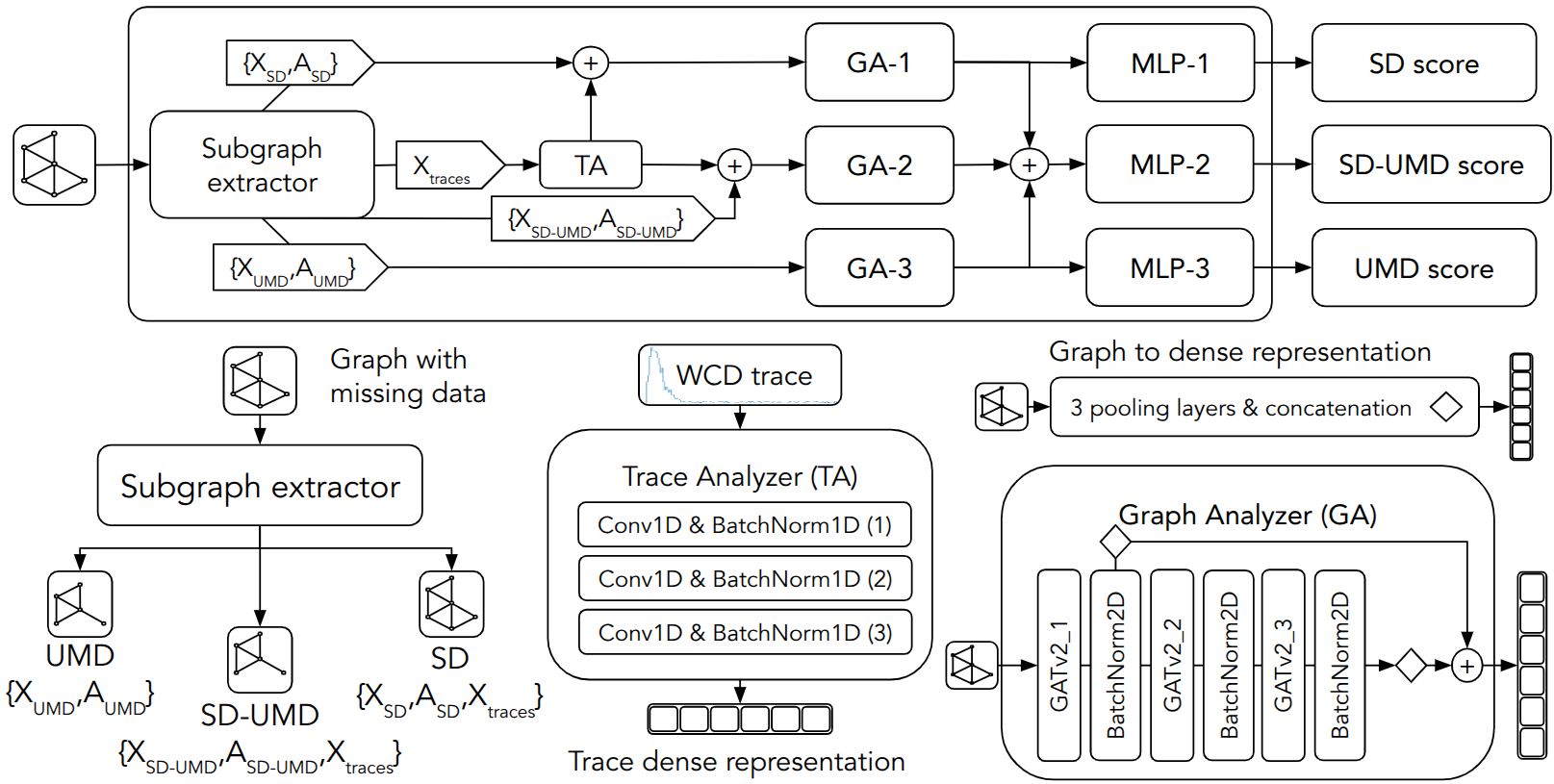}
    \caption{Schematic of the updated neural network architecture and its modular subcomponents. A single graph encodes SD and UMD information. Binary masks guide the Subgraph Extractor in generating SD, UMD, and SD–UMD subgraphs. Trace features are computed and added to the relevant nodes before graph attention layers (GATv2) are applied. Global pooling operations (mean, max, and sum) are performed on the first and third GATv2 outputs, and their concatenation is passed to MLPs to obtain photon-like scores.}
    \label{fig:GNN}
\end{figure}

\cref{fig:GNN} shows a sketch of the improved architecture. The \emph{Trace Analyzer} (TA), unchanged from the previous network version, processes SD time traces using 1D convolutional layers, producing features concatenated with the node attributes for SD and SD–UMD processing. The node features are processed through three distinct \emph{Graph Analyzers} (GA) corresponding to SD, UMD, and SD–UMD. Each GA consists of three GATv2 layers~\cite{Brody2022}. In the updated network, pooling operations (mean, max, sum) are applied to both the first and third outputs of the GATv2 layers, and the results are concatenated before being passed to the classification multi-layer perceptrons (MLPs). This enriches the final graph embedding by combining low- and high-level representations. Each of the three graph-level embeddings is then processed by a dedicated MLP to produce a photon-like classification score.

Unlike the earlier joint training strategy, the network is currently trained using a sequential and independent path optimization approach: the UMD path is trained first; all other weights, including the Trace Analyzer, other GATv2 layers, and MLPs, are kept frozen; next, the SD path is trained, with its corresponding GATv2 and MLP, as well as the shared Trace Analyzer, being updated while the rest of the network remains fixed; finally, the SD–UMD path is trained using its dedicated GATv2 and MLP, keeping all previously trained components frozen.

This training strategy allows each head to specialize in extracting the most informative features from its respective data stream, while avoiding inter-path interference. Each sub-network is optimized using the binary cross-entropy loss between its photon-like score and the true event label.

\begin{figure}
    \centering
    \includegraphics[width=\textwidth]{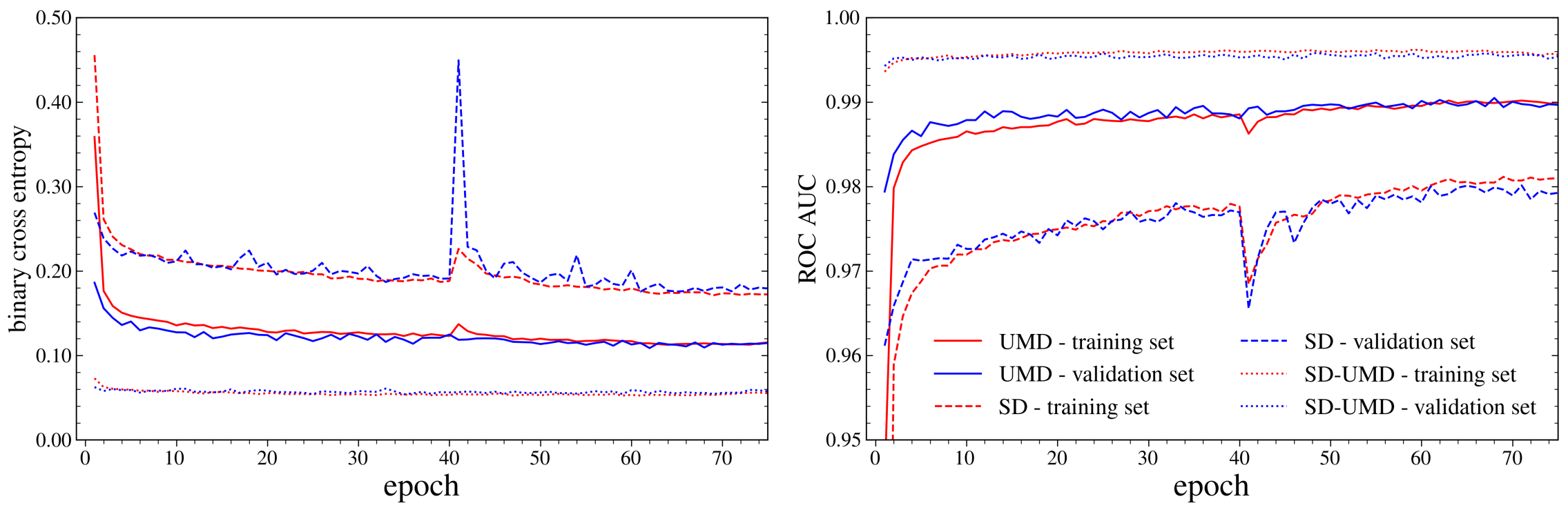}
    \caption{Training and validation performance for each path. Left panel: binary cross-entropy loss. Right panel: ROC-AUC scores. The close match between training and validation curves indicates stable generalization.}
    \label{fig:learning_cruves}
\end{figure}

As shown in~\cref{fig:learning_cruves}, the training curves for each information path remain closely aligned with their respective validation metrics, indicating no significant overfitting. Among the three outputs, the SD–UMD path consistently exhibits the highest performance. Notably, the UMD-only path also provides strong discrimination, reflecting the value of direct muon measurements in photon searches.

\section{Photon-hadron discrimination}
\label{sec:discrimination}

\begin{wrapfigure}{r}{0.5\textwidth}
    \centering
    \vspace{-10pt}
    \includegraphics[width=0.5\textwidth]{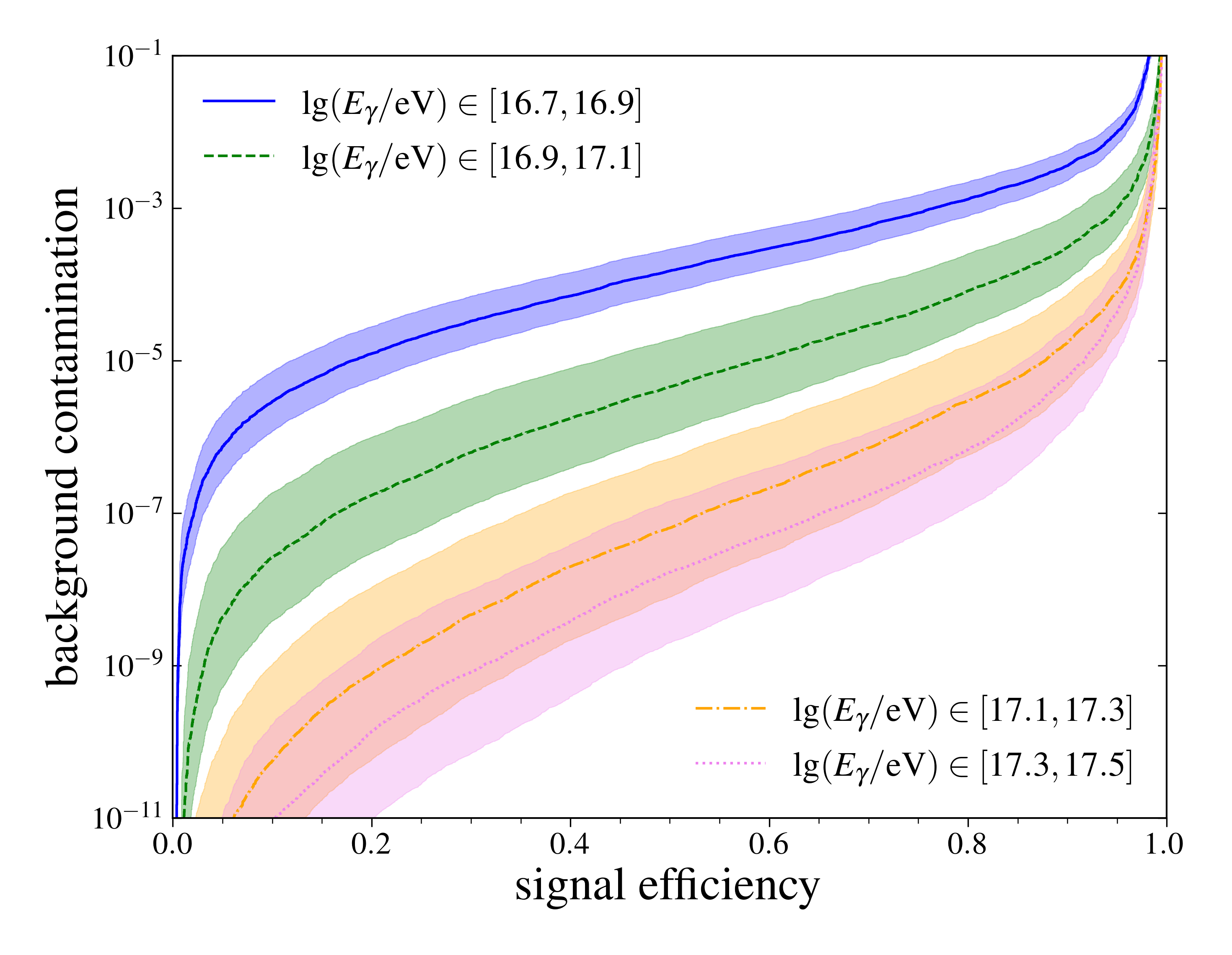}
    \caption{Estimated background contamination vs.\ signal efficiency in four energy intervals. Shaded bands correspond to bootstrap uncertainties on the exponential tail fits used to model the background score distributions.}
    \label{fig:roc}
    \vspace{-10pt}
\end{wrapfigure}

The trained deep learning model provides effective separation between photon and hadron primaries, with the SD–UMD score consistently outperforming the UMD-only and SD-only classifiers. This ordering in discrimination power is already observed in the training and validation curves (see~\cref{fig:learning_cruves}).

To quantify background contamination, we focus on the SD–UMD score as the most powerful discriminator. Figure~\ref{fig:roc} shows the estimated background contamination as a function of signal efficiency in four energy bins. Contamination was computed by modeling the score distribution of the most photon-like proton events using exponential fits to the tail region. Due to the limited number of events in the tails, a bootstrapping procedure was employed to estimate the uncertainty on the slope. Small variations in the slope significantly affect the background contamination expectation at high rejection levels.

At a fixed signal efficiency of 0.5, the average background contamination in the bins $\lg(E_{\upgamma}/\mathrm{eV}) \in [16.7, 16.9)$, $[16.9, 17.1)$, $[17.1, 17.3)$, and $[17.3, 17.5)$ is below $8 \times 10^{-5}$, $10^{-5}$, $10^{-7}$, and $10^{-8}$, respectively. It is worth noting that these estimates include the effect of random PMT and UMD module masking, as well as entire SD or UMD station removal, to emulate real operation conditions. As such, they represent conservative values compared to idealized simulations.

To validate the applicability of the model to experimental data, we use a subset of events that were previously burnt in a photon search using the SD-433 and UMD arrays~\cite{JCAP2025PhotonHadron}. These events were acquired between December 2020 and March 2022 and correspond to approximately 10\% of the complete high-quality dataset. 

\begin{figure}[htbp]
    \centering
    \includegraphics[width=\textwidth]{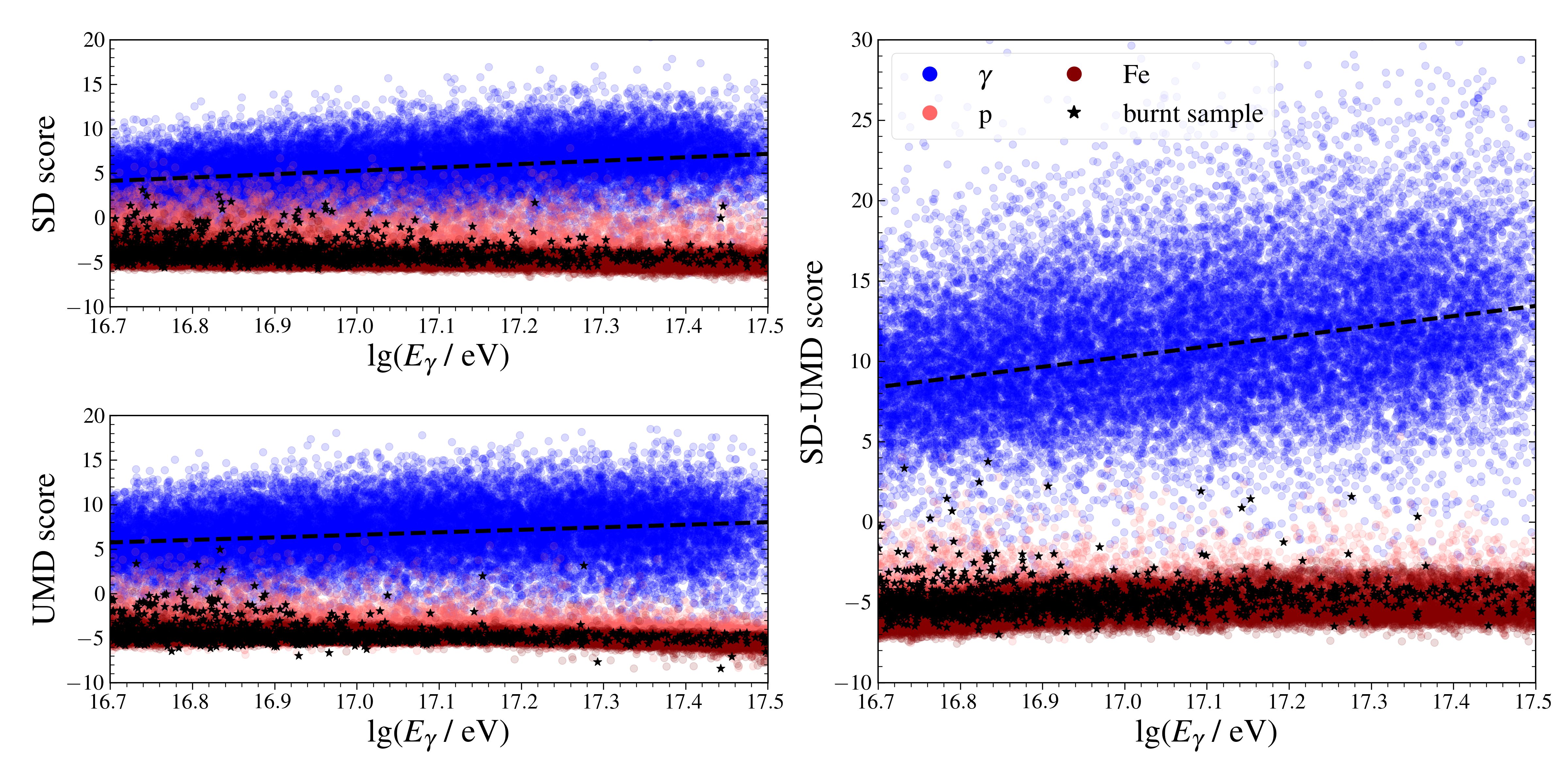}
    \caption{Classification scores from the SD (top left), UMD (bottom left), and SD–UMD (right) heads as a function of photon energy. Simulated photons, protons, and iron nuclei are shown, along with burnt sample events (black stars). Dashed lines indicate candidate selection thresholds based on the photon medians.}
    \label{fig:scores}
\end{figure}

For photon search applications, a candidate selection cut must be defined. Following standard practice, we used the median of the photon score distributions in each energy bin to set the candidate threshold. A linear model is used to describe the energy dependence. Figure~\ref{fig:scores} exhibits the scores from all three classifiers as a function of the photon energy. Simulated photons, protons, and iron nuclei are displayed for comparison, along with the burnt sample (black stars).

All three scores yield distributions that clearly separate photons from hadronic backgrounds, with the SD–UMD score providing the cleanest separation. None of the events from the burnt dataset cross the candidate cut in any of the three scores. The absence of candidates is consistent with a fully hadronic population and aligns with the findings reported in~\cite{JCAP2025PhotonHadron}.

\section{Summary and outlook}
\label{summary}

We have developed a graph-based neural network to discriminate between photon and hadron primaries in extensive air showers measured at the Pierre Auger Observatory. By encoding the spatio-temporal information of the SD-433 and the UMD into a unified graph representation, and processing these through a multi-path architecture with graph attention layers, the model achieves strong photon–hadron separation across the energy range $50$–$300$\,PeV. Using the SD-UMD score of the GNN we achieve a background contamination level of at least $10^{-5}$ for a signal efficiency of 0.5 in realistic scenarios where one or more UMD stations are not operative.

The model was trained and validated on a realistic simulation data set that includes detector inefficiencies and reconstruction with photon priors. Its performance was assessed on both test simulations and a previously burnt subset of data, showing compatibility with hadronic expectations and validating the application of deep learning techniques to data. Conservative background contamination estimates were derived through exponential tail fits with bootstrap uncertainties, which underscore the current limitation posed by the low statistics in the photon-like tail of proton events.

Deep learning methods now complement traditional photon search techniques at the Pierre Auger Observatory. As with previous approaches, the main limitation remains to be the exposure, due to the extremely low expected photon flux. With accumulating statistics, future efforts should prioritize more accurate background modeling and the enhancement of the sensitivity to photon primaries. Whether through a discovery or the setting of tighter upper limits, these searches will increasingly constrain cosmogenic photon fluxes and models involving SHDM particles. In particular, the growing exposure of the SD-433 array and UMD will allow us to test the hypothesis of a diffuse photon flux originating from \emph{pp} interactions in the Galactic halo~\cite{Kalashev2014}, as well as predictions from two distinct SHDM decay scenarios~\cite{Kalashev2016,Kachelriess2018}.

\clearpage

\section*{The Pierre Auger Collaboration}

{\footnotesize\setlength{\baselineskip}{10pt}
\noindent
\begin{wrapfigure}[11]{l}{0.12\linewidth}
\vspace{-4pt}
\includegraphics[width=0.98\linewidth]{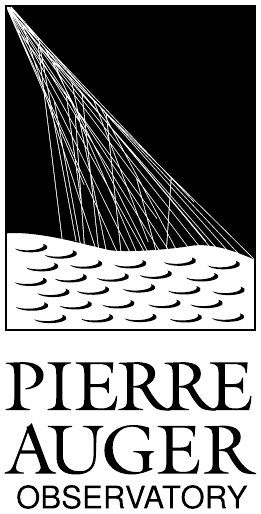}
\end{wrapfigure}
\begin{sloppypar}\noindent
\input{latex_authorlist_authors}
\end{sloppypar}
\begin{center}
\end{center}

\vspace{1ex}
\input{latex_authorlist_institutions}

\clearpage
\input{acknowledgments}
}

\end{document}

%% file: latex_authorlist_authors.tex
A.~Abdul Halim$^{13}$,
P.~Abreu$^{70}$,
M.~Aglietta$^{53,51}$,
I.~Allekotte$^{1}$,
K.~Almeida Cheminant$^{78,77}$,
A.~Almela$^{7,12}$,
R.~Aloisio$^{44,45}$,
J.~Alvarez-Mu\~niz$^{76}$,
A.~Ambrosone$^{44}$,
J.~Ammerman Yebra$^{76}$
% ...

%% file: latex_authorlist_institutions.tex
\begin{description}[labelsep=0.2em,align=right,labelwidth=0.7em,labelindent=0em,leftmargin=2em,noitemsep,before={\renewcommand\makelabel[1]{##1 }}]
\item[$^{1}$] Centro At\'omico Bariloche and Instituto Balseiro (CNEA-UNCuyo-CONICET), San Carlos de Bariloche, Argentina
\item[$^{2}$] Departamento de F\'\i{}sica and Departamento de Ciencias de la Atm\'osfera y los Oc\'eanos, FCEyN, Universidad de Buenos Aires and CONICET, Buenos Aires, Argentina
\item[$^{3}$] IFLP, Universidad Nacional de La Plata and CONICET, La Plata, Argentina
\item[$^{4}$] Instituto de Astronom\'\i{}a y F\'\i{}sica del Espacio (IAFE, CONICET-UBA), Buenos Aires, Argentina
\item[$^{5}$] Instituto de F\'\i{}sica de Rosario (IFIR) -- CONICET/U.N.R.\ and Facultad de Ciencias Bioqu\'\i{}micas y Farmac\'euticas U.N.R., Rosario, Argentina
\item[$^{6}$] Instituto de Tecnolog\'\i{}as en Detecci\'on y Astropart\'\i{}culas (CNEA, CONICET, UNSAM), and Universidad Tecnol\'ogica Nacional -- Facultad Regional Mendoza (CONICET/CNEA), Mendoza, Argentina
\item[$^{7}$] Instituto de Tecnolog\'\i{}as en Detecci\'on y Astropart\'\i{}culas (CNEA, CONICET, UNSAM), Buenos Aires, Argentina
\item[$^{8}$] International Center of Advanced Studies and Instituto de Ciencias F\'\i{}sicas, ECyT-UNSAM and CONICET, Campus Miguelete -- San Mart\'\i{}n, Buenos Aires, Argentina
\item[$^{9}$] Laboratorio Atm\'osfera -- Departamento de Investigaciones en L\'aseres y sus Aplicaciones -- UNIDEF (CITEDEF-CONICET), Argentina
\item[$^{10}$] Observatorio Pierre Auger, Malarg\"ue, Argentina
\item[$^{11}$] Observatorio Pierre Auger and Comisi\'on Nacional de Energ\'\i{}a At\'omica, Malarg\"ue, Argentina
\item[$^{12}$] Universidad Tecnol\'ogica Nacional -- Facultad Regional Buenos Aires, Buenos Aires, Argentina
\item[$^{13}$] University of Adelaide, Adelaide, S.A., Australia
\item[$^{14}$] Universit\'e Libre de Bruxelles (ULB), Brussels, Belgium
\item[$^{15}$] Vrije Universiteit Brussels, Brussels, Belgium
\item[$^{16}$] Centro Brasileiro de Pesquisas Fisicas, Rio de Janeiro, RJ, Brazil
\item[$^{17}$] Centro Federal de Educa\c{c}\~ao Tecnol\'ogica Celso Suckow da Fonseca, Petropolis, Brazil
\item[$^{18}$] Instituto Federal de Educa\c{c}\~ao, Ci\^encia e Tecnologia do Rio de Janeiro (IFRJ), Brazil
\item[$^{19}$] Universidade de S\~ao Paulo, Escola de Engenharia de Lorena, Lorena, SP, Brazil
\item[$^{20}$] Universidade de S\~ao Paulo, Instituto de F\'\i{}sica de S\~ao Carlos, S\~ao Carlos, SP, Brazil
\item[$^{21}$] Universidade de S\~ao Paulo, Instituto de F\'\i{}sica, S\~ao Paulo, SP, Brazil
\item[$^{22}$] Universidade Estadual de Campinas (UNICAMP), IFGW, Campinas, SP, Brazil
\item[$^{23}$] Universidade Estadual de Feira de Santana, Feira de Santana, Brazil
\item[$^{24}$] Universidade Federal de Campina Grande, Centro de Ciencias e Tecnologia, Campina Grande, Brazil
\item[$^{25}$] Universidade Federal do ABC, Santo Andr\'e, SP, Brazil
\item[$^{26}$] Universidade Federal do Paran\'a, Setor Palotina, Palotina, Brazil
\item[$^{27}$] Universidade Federal do Rio de Janeiro, Instituto de F\'\i{}sica, Rio de Janeiro, RJ, Brazil
\item[$^{28}$] Universidad de Medell\'\i{}n, Medell\'\i{}n, Colombia
\item[$^{29}$] Universidad Industrial de Santander, Bucaramanga, Colombia
\item[$^{30}$] Charles University, Faculty of Mathematics and Physics, Institute of Particle and Nuclear Physics, Prague, Czech Republic
\item[$^{31}$] Institute of Physics of the Czech Academy of Sciences, Prague, Czech Republic
\item[$^{32}$] Palacky University, Olomouc, Czech Republic
\item[$^{33}$] CNRS/IN2P3, IJCLab, Universit\'e Paris-Saclay, Orsay, France
\item[$^{34}$] Laboratoire de Physique Nucl\'eaire et de Hautes Energies (LPNHE), Sorbonne Universit\'e, Universit\'e de Paris, CNRS-IN2P3, Paris, France
\item[$^{35}$] Univ.\ Grenoble Alpes, CNRS, Grenoble Institute of Engineering Univ.\ Grenoble Alpes, LPSC-IN2P3, 38000 Grenoble, France
\item[$^{36}$] Universit\'e Paris-Saclay, CNRS/IN2P3, IJCLab, Orsay, France
\item[$^{37}$] Bergische Universit\"at Wuppertal, Department of Physics, Wuppertal, Germany
\item[$^{38}$] Karlsruhe Institute of Technology (KIT), Institute for Experimental Particle Physics, Karlsruhe, Germany
\item[$^{39}$] Karlsruhe Institute of Technology (KIT), Institut f\"ur Prozessdatenverarbeitung und Elektronik, Karlsruhe, Germany
\item[$^{40}$] Karlsruhe Institute of Technology (KIT), Institute for Astroparticle Physics, Karlsruhe, Germany
\item[$^{41}$] RWTH Aachen University, III.\ Physikalisches Institut A, Aachen, Germany
\item[$^{42}$] Universit\"at Hamburg, II.\ Institut f\"ur Theoretische Physik, Hamburg, Germany
\item[$^{43}$] Universit\"at Siegen, Department Physik -- Experimentelle Teilchenphysik, Siegen, Germany
\item[$^{44}$] Gran Sasso Science Institute, L'Aquila, Italy
\item[$^{45}$] INFN Laboratori Nazionali del Gran Sasso, Assergi (L'Aquila), Italy
\item[$^{46}$] INFN, Sezione di Catania, Catania, Italy
\item[$^{47}$] INFN, Sezione di Lecce, Lecce, Italy
\item[$^{48}$] INFN, Sezione di Milano, Milano, Italy
\item[$^{49}$] INFN, Sezione di Napoli, Napoli, Italy
\item[$^{50}$] INFN, Sezione di Roma ``Tor Vergata'', Roma, Italy
\item[$^{51}$] INFN, Sezione di Torino, Torino, Italy
\item[$^{52}$] Istituto di Astrofisica Spaziale e Fisica Cosmica di Palermo (INAF), Palermo, Italy
\item[$^{53}$] Osservatorio Astrofisico di Torino (INAF), Torino, Italy
\item[$^{54}$] Politecnico di Milano, Dipartimento di Scienze e Tecnologie Aerospaziali , Milano, Italy
\item[$^{55}$] Universit\`a del Salento, Dipartimento di Matematica e Fisica ``E.\ De Giorgi'', Lecce, Italy
\item[$^{56}$] Universit\`a dell'Aquila, Dipartimento di Scienze Fisiche e Chimiche, L'Aquila, Italy
\item[$^{57}$] Universit\`a di Catania, Dipartimento di Fisica e Astronomia ``Ettore Majorana``, Catania, Italy
\item[$^{58}$] Universit\`a di Milano, Dipartimento di Fisica, Milano, Italy
\item[$^{59}$] Universit\`a di Napoli ``Federico II'', Dipartimento di Fisica ``Ettore Pancini'', Napoli, Italy
\item[$^{60}$] Universit\`a di Palermo, Dipartimento di Fisica e Chimica ''E.\ Segr\`e'', Palermo, Italy
\item[$^{61}$] Universit\`a di Roma ``Tor Vergata'', Dipartimento di Fisica, Roma, Italy
\item[$^{62}$] Universit\`a Torino, Dipartimento di Fisica, Torino, Italy
\item[$^{63}$] Benem\'erita Universidad Aut\'onoma de Puebla, Puebla, M\'exico
\item[$^{64}$] Unidad Profesional Interdisciplinaria en Ingenier\'\i{}a y Tecnolog\'\i{}as Avanzadas del Instituto Polit\'ecnico Nacional (UPIITA-IPN), M\'exico, D.F., M\'exico
\item[$^{65}$] Universidad Aut\'onoma de Chiapas, Tuxtla Guti\'errez, Chiapas, M\'exico
\item[$^{66}$] Universidad Michoacana de San Nicol\'as de Hidalgo, Morelia, Michoac\'an, M\'exico
\item[$^{67}$] Universidad Nacional Aut\'onoma de M\'exico, M\'exico, D.F., M\'exico
\item[$^{68}$] Institute of Nuclear Physics PAN, Krakow, Poland
\item[$^{69}$] University of \L{}\'od\'z, Faculty of High-Energy Astrophysics,\L{}\'od\'z, Poland
\item[$^{70}$] Laborat\'orio de Instrumenta\c{c}\~ao e F\'\i{}sica Experimental de Part\'\i{}culas -- LIP and Instituto Superior T\'ecnico -- IST, Universidade de Lisboa -- UL, Lisboa, Portugal
\item[$^{71}$] ``Horia Hulubei'' National Institute for Physics and Nuclear Engineering, Bucharest-Magurele, Romania
\item[$^{72}$] Institute of Space Science, Bucharest-Magurele, Romania
\item[$^{73}$] Center for Astrophysics and Cosmology (CAC), University of Nova Gorica, Nova Gorica, Slovenia
\item[$^{74}$] Experimental Particle Physics Department, J.\ Stefan Institute, Ljubljana, Slovenia
\item[$^{75}$] Universidad de Granada and C.A.F.P.E., Granada, Spain
\item[$^{76}$] Instituto Galego de F\'\i{}sica de Altas Enerx\'\i{}as (IGFAE), Universidade de Santiago de Compostela, Santiago de Compostela, Spain
\item[$^{77}$] IMAPP, Radboud University Nijmegen, Nijmegen, The Netherlands
\item[$^{78}$] Nationaal Instituut voor Kernfysica en Hoge Energie Fysica (NIKHEF), Science Park, Amsterdam, The Netherlands
\item[$^{79}$] Stichting Astronomisch Onderzoek in Nederland (ASTRON), Dwingeloo, The Netherlands
\item[$^{80}$] Universiteit van Amsterdam, Faculty of Science, Amsterdam, The Netherlands
\item[$^{81}$] Case Western Reserve University, Cleveland, OH, USA
\item[$^{82}$] Colorado School of Mines, Golden, CO, USA
\item[$^{83}$] Department of Physics and Astronomy, Lehman College, City University of New York, Bronx, NY, USA
\item[$^{84}$] Michigan Technological University, Houghton, MI, USA
\item[$^{85}$] New York University, New York, NY, USA
\item[$^{86}$] University of Chicago, Enrico Fermi Institute, Chicago, IL, USA
\item[$^{87}$] University of Delaware, Department of Physics and Astronomy, Bartol Research Institute, Newark, DE, USA
\item[] -----
\item[$^{a}$] Max-Planck-Institut f\"ur Radioastronomie, Bonn, Germany
\item[$^{b}$] also at Kapteyn Institute, University of Groningen, Groningen, The Netherlands
\item[$^{c}$] School of Physics and Astronomy, University of Leeds, Leeds, United Kingdom
\item[$^{d}$] Fermi National Accelerator Laboratory, Fermilab, Batavia, IL, USA
\item[$^{e}$] Pennsylvania State University, University Park, PA, USA
\item[$^{f}$] Colorado State University, Fort Collins, CO, USA
\item[$^{g}$] Louisiana State University, Baton Rouge, LA, USA
\item[$^{h}$] now at Graduate School of Science, Osaka Metropolitan University, Osaka, Japan
\item[$^{i}$] Institut universitaire de France (IUF), France
\item[$^{j}$] now at Technische Universit\"at Dortmund and Ruhr-Universit\"at Bochum, Dortmund and Bochum, Germany
\end{description}

%% file: acknowledgments.tex
\section*{Acknowledgments}

\begin{sloppypar}
The successful installation, commissioning, and operation of the Pierre
Auger Observatory would not have been possible without the strong
commitment and effort from the technical and administrative staff in
Malarg\"ue. We are very grateful to the following agencies and
organizations for financial support:
\end{sloppypar}

\begin{sloppypar}
Argentina -- Comisi\'on Nacional de Energ\'\i{}a At\'omica; Agencia Nacional de
Promoci\'on Cient\'\i{}fica y Tecnol\'ogica (ANPCyT); Consejo Nacional de
Investigaciones Cient\'\i{}ficas y T\'ecnicas (CONICET); Gobierno de la
Provincia de Mendoza; Municipalidad de Malarg\"ue; NDM Holdings and Valle
Las Le\~nas; in gratitude for their continuing cooperation over land
access; Australia -- the Australian Research Council; Belgium -- Fonds
de la Recherche Scientifique (FNRS); Research Foundation Flanders (FWO),
Marie Curie Action of the European Union Grant No.~101107047; Brazil --
Conselho Nacional de Desenvolvimento Cient\'\i{}fico e Tecnol\'ogico (CNPq);
Financiadora de Estudos e Projetos (FINEP); Funda\c{c}\~ao de Amparo \`a
Pesquisa do Estado de Rio de Janeiro (FAPERJ); S\~ao Paulo Research
Foundation (FAPESP) Grants No.~2019/10151-2, No.~2010/07359-6 and
No.~1999/05404-3; Minist\'erio da Ci\^encia, Tecnologia, Inova\c{c}\~oes e
Comunica\c{c}\~oes (MCTIC); Czech Republic -- GACR 24-13049S, CAS LQ100102401,
MEYS LM2023032, CZ.02.1.01/0.0/0.0/16{\textunderscore}013/0001402,
CZ.02.1.01/0.0/0.0/18{\textunderscore}046/0016010 and
CZ.02.1.01/0.0/0.0/17{\textunderscore}049/0008422 and CZ.02.01.01/00/22{\textunderscore}008/0004632;
France -- Centre de Calcul IN2P3/CNRS; Centre National de la Recherche
Scientifique (CNRS); Conseil R\'egional Ile-de-France; D\'epartement
Physique Nucl\'eaire et Corpusculaire (PNC-IN2P3/CNRS); D\'epartement
Sciences de l'Univers (SDU-INSU/CNRS); Institut Lagrange de Paris (ILP)
Grant No.~LABEX ANR-10-LABX-63 within the Investissements d'Avenir
Programme Grant No.~ANR-11-IDEX-0004-02; Germany -- Bundesministerium
f\"ur Bildung und Forschung (BMBF); Deutsche Forschungsgemeinschaft (DFG);
Finanzministerium Baden-W\"urttemberg; Helmholtz Alliance for
Astroparticle Physics (HAP); Helmholtz-Gemeinschaft Deutscher
Forschungszentren (HGF); Ministerium f\"ur Kultur und Wissenschaft des
Landes Nordrhein-Westfalen; Ministerium f\"ur Wissenschaft, Forschung und
Kunst des Landes Baden-W\"urttemberg; Italy -- Istituto Nazionale di
Fisica Nucleare (INFN); Istituto Nazionale di Astrofisica (INAF);
Ministero dell'Universit\`a e della Ricerca (MUR); CETEMPS Center of
Excellence; Ministero degli Affari Esteri (MAE), ICSC Centro Nazionale
di Ricerca in High Performance Computing, Big Data and Quantum
Computing, funded by European Union NextGenerationEU, reference code
CN{\textunderscore}00000013; M\'exico -- Consejo Nacional de Ciencia y Tecnolog\'\i{}a
(CONACYT) No.~167733; Universidad Nacional Aut\'onoma de M\'exico (UNAM);
PAPIIT DGAPA-UNAM; The Netherlands -- Ministry of Education, Culture and
Science; Netherlands Organisation for Scientific Research (NWO); Dutch
national e-infrastructure with the support of SURF Cooperative; Poland
-- Ministry of Education and Science, grants No.~DIR/WK/2018/11 and
2022/WK/12; National Science Centre, grants No.~2016/22/M/ST9/00198,
2016/23/B/ST9/01635, 2020/39/B/ST9/01398, and 2022/45/B/ST9/02163;
Portugal -- Portuguese national funds and FEDER funds within Programa
Operacional Factores de Competitividade through Funda\c{c}\~ao para a Ci\^encia
e a Tecnologia (COMPETE); Romania -- Ministry of Research, Innovation
and Digitization, CNCS-UEFISCDI, contract no.~30N/2023 under Romanian
National Core Program LAPLAS VII, grant no.~PN 23 21 01 02 and project
number PN-III-P1-1.1-TE-2021-0924/TE57/2022, within PNCDI III; Slovenia
-- Slovenian Research Agency, grants P1-0031, P1-0385, I0-0033, N1-0111;
Spain -- Ministerio de Ciencia e Innovaci\'on/Agencia Estatal de
Investigaci\'on (PID2019-105544GB-I00, PID2022-140510NB-I00 and
RYC2019-027017-I), Xunta de Galicia (CIGUS Network of Research Centers,
Consolidaci\'on 2021 GRC GI-2033, ED431C-2021/22 and ED431F-2022/15),
Junta de Andaluc\'\i{}a (SOMM17/6104/UGR and P18-FR-4314), and the European
Union (Marie Sklodowska-Curie 101065027 and ERDF); USA -- Department of
Energy, Contracts No.~DE-AC02-07CH11359, No.~DE-FR02-04ER41300,
No.~DE-FG02-99ER41107 and No.~DE-SC0011689; National Science Foundation,
Grant No.~0450696, and NSF-2013199; The Grainger Foundation; Marie
Curie-IRSES/EPLANET; European Particle Physics Latin American Network;
and UNESCO.
\end{sloppypar}